\documentclass{aa}  

\usepackage{graphicx}
\usepackage{subcaption}
\usepackage{physics}
\usepackage{xcolor}
\usepackage{float}
\usepackage[normalem]{ulem}
\usepackage[switch]{lineno}
\usepackage{amsmath}
\usepackage{comment}
\usepackage{placeins}
\usepackage{txfonts}
\usepackage[hidelinks]{hyperref}
\usepackage{cleveref}
\crefname{figure}{Fig.}{Figs.}

\newcommand{\ha}{{H$\alpha$ }}

\newcommand{\ML}[2]{\sout{#1}{\bf \textcolor{olive}{#2}}}

\begin{document}

\title{Automatic detection of solar filament oscillations I: Multi-scale spectral pipeline}
   \subtitle{}

\author{G. Castelló
      \inst{1,2}
      \and
      M. Luna 
      \inst{1,2}
      \and
      J. Terradas
      \inst{1,2}
      }

\institute{Departament de Física, Universitat de les Illes Balears (UIB), E-07122 Palma de Mallorca, Spain\\
          \email{guillem.castello@uib.cat}
     \and
         Institute of Applied Computing \& Community Code (IAC3), UIB, E-07122 Palma de Mallorca, Spain}

   \date{Received May 05, 2026; accepted July 01, 2026}

   \abstract
{Solar filament oscillations are valuable probes of the magnetic structure and stability of prominences. Although long-duration, high-cadence H$\alpha$ archives such as those from GONG make systematic studies possible, most oscillation detections still rely on visual inspection, manually placed slits, and time--distance diagrams. These methods are difficult to scale and are biased toward the most conspicuous events.}
{We present and validate an automatic pipeline for detecting spatially coherent oscillations in solar filaments observed by GONG. The goal is to recover robust filament-scale oscillatory signals while suppressing local intensity fluctuations that dominate pixel-wise spectral analyses.}
{Daily GONG H$\alpha$ image sequences are preprocessed, coaligned, and corrected for large-scale intensity variations. Filaments are identified using deep-learning detection and segmentation models. For each filament region, we perform a multi-scale analysis in which the image sequence is spatially averaged over overlapping kernels of different sizes. Lomb--Scargle periodograms are computed for the resulting time series, the spectral background is estimated with a pre-trained convolutional neural network, and significance thresholds are obtained through an empirical conformal-style calibration. Candidate detections are then grouped in period and space, and only events supported by at least four spatial scales are retained.}
{The pipeline successfully recovers several events reported in the manual GONG catalog of \citet{luna_gong_2018}, including the 1 January 2014 oscillation with a period of about 76 min. Applied to the first two weeks of January 2014, the method identifies 91 oscillatory events, compared with 22 events in the corresponding manual catalog after duplicate removal. The detected events have periods between approximately 20 and 126 min. The pipeline also reveals oscillations on days for which no event was previously reported; one such case on 13 January 2014, with a period of about 86 min, is independently confirmed using a conventional time--distance diagram.}
{The proposed framework provides a reproducible and scalable alternative to manual slit-based searches. By combining automated filament segmentation, multi-scale spectral analysis, and calibrated significance thresholds, it preferentially selects coherent filament-scale oscillations and substantially increases detection sensitivity. This method establishes the basis for future statistical studies of filament oscillations over extended GONG time intervals and, ultimately, across the solar cycle.}


\maketitle
\nolinenumbers
%
\section{Introduction}
Solar prominences are cool, dense plasma structures suspended in the hot corona by magnetic fields. When observed on the solar disk they appear in absorption and are commonly referred to as filaments. Because their morphology and dynamics are tightly linked to the underlying magnetic configuration, filaments provide a valuable diagnostic of the large-scale solar atmosphere and of the processes that precede and accompany eruptive activity \citep{Mackay2010, Gibson2018}. Among their many dynamical phenomena, oscillatory motions are of particular interest because they offer a way to infer physical properties of the supporting magnetic structure that are otherwise difficult to measure directly.

Oscillations in prominences and filaments have been reported for decades and over a broad range of periods, amplitudes, and spatial scales. Early reviews already established that these motions can be interpreted in terms of magnetohydrodynamic waves and can be exploited for prominence seismology \citep{OliverBallester2002}. Later works showed that gravity is also relevant in longitudinal oscillations, the so-called gravitoacoustic modes \citep[i.e.][]{luna_large-amplitude_2012,zhang_observations_2012}.
These works demonstrated that both transverse and longitudinal oscillations encode information on magnetic-field strength, field-line geometry, and damping processes \citep{Tripathi2009, arregui_prominence_2018}. In this context, filament oscillations are not only a manifestation of solar dynamics but also a diagnostic tool for probing prominence structure and stability.

A substantial fraction of the observational literature has relied on case studies in which the oscillation is identified visually and analyzed through manually placed slits and time--distance diagrams. This approach has proved very successful for detailed characterization of individual events, including oscillations triggered by subflares, nearby flares, jets, Moreton waves, and EUV waves \citep{okamoto2004,jing2006,vrsnak2007,shen2014,zhang2017, luna_gong_2018,luna2024,beckwith2024,gao2025}. These studies have demonstrated the diversity of filament oscillations and their triggering mechanisms, while also reinforcing the value of seismological interpretations. At the same time, however, slit-based analyses are inherently labor-intensive, depend on prior visual identification of the event, and are difficult to scale to the large observational archives that are now available.

Observations have also revealed that filament oscillations span a wide range of characteristic timescales. In addition to the more commonly reported periods of tens of minutes to a few hours, long and ultra-long-period oscillations have been detected in both EUV and H$\alpha$ observations, extending the phenomenology to timescales of many hours \citep{foullon_detection_2004,foullon_ultra-long-period_2009,efremov_ultra_2016}. This broad dynamical range reinforces the need for analysis methods that are flexible with respect to temporal sampling and capable of handling long, heterogeneous time series.

The Global Oscillation Network Group (GONG) is particularly well suited to this task because it provides near-continuous full-disk H$\alpha$ monitoring, in which filaments are seen with strong contrast against the chromospheric background. Using these data, \citet{luna_gong_2018} produced the first GONG catalog of solar filament oscillations near solar maximum, demonstrating the scientific potential of systematic searches in long H$\alpha$ time series. More recently, \citet{luna2022} introduced an automatic proof-of-concept spectral technique for detecting filament oscillations in GONG data, and \citet{castello2025} showed that convolutional neural networks can accelerate the estimation of spectral backgrounds and significance levels while retaining agreement with more expensive Bayesian approaches. These developments mark an important transition from purely manual analyses towards automated, large-scale studies.

Nevertheless, significant challenges remain. Pixel-wise spectral analysis often generates an overwhelming volume of frequency information where local fluctuations mask the collective motion of the filament. This results in a fragmented spatial distribution of signals, appearing as disjointed patches rather than a coherent global mode. This high degree of dispersion in both frequency and space significantly limits the method’s utility for identifying large-scale oscillations.

Furthermore, the neural network approach introduced in \citet{castello2025} is notably rigid; its fixed architecture does not allow for the adjustment of confidence levels, making it unable to adapt to the variable quality of ground-based data\footnote{We showed a model capable of predicting the 95\% confidence line for a given \ha power spectral density, but we empirically validated such level to be too small for our research goals.}. A practical automatic method must therefore do more than detect significant peaks: it must identify oscillatory signals that are statistically reliable and representative of the filament as a whole.

In this paper, to address these limitations, we present a novel automatic pipeline for the detection of solar filament oscillations in GONG H$\alpha$ observations. 
Our approach bridges the gap between local and global scales by combining filament detection and segmentation with a multi-scale spectral analysis designed to emphasize coherent motions over extended filament regions.
The spectral background is estimated using a pre-trained neural network, while statistical significance is calibrated through a conformal-style empirical procedure. 
Candidate detections are then consolidated across period and space in order to recover robust oscillatory events. The goal of this work is to establish this framework as a systematic solution for filament studies, detailing its design and validating its performance against both previously reported events and the GONG catalog of \citet{luna_gong_2018}.

The paper is organized as follows. In Sect.~2 we describe the data set, preprocessing steps, and filament detection and segmentation. In Sect.~3 we present the automatic oscillation-analysis pipeline, including the multi-scale strategy and the conformal calibration procedure. In Sect.~4 we show representative results and compare the detected events with those reported by \citet{luna_gong_2018}. Finally, Sect.~5 summarizes the main conclusions.

\section{Data}
In \ha observations, solar filaments are visible as dark, elongated structures set against the bright chromospheric background, making this wavelength particularly suitable for investigating filament dynamics. Such clarity is generally not present in extreme ultraviolet (EUV) data obtained from instruments like the Atmospheric Imaging Assembly (AIA) aboard the Solar Dynamics Observatory \citep[SDO;][]{lemen_atmospheric_2011}, where the presence of more intricate and rapidly evolving coronal features, often acting as a bright foreground that masks the filaments, complicates both the identification of filaments and the automated detection of their oscillatory behavior.

\subsection{NSO-GONG network}
The NSO-GONG \footnote{\url{http://gong2.nso.edu}} \citep{hill_global_1994} telescope network provides nearly continuous full-disk monitoring of the Sun in the \ha wavelength.
The network is composed of telescopes with identical design and instrumentation, strategically distributed around the globe at six locations: Learmonth (L), Udaipur (U), El Teide (T), Cerro Tololo (C), Big Bear (B), and Mauna Loa (M). This configuration was specifically chosen to follow the Sun’s diurnal motion and to ensure uninterrupted daily coverage.
Each station acquires images during its local daytime, with consecutive sites providing overlapping observation periods as the Earth rotates.

Because the NSO-GONG network operates from ground-based observatories, variations in atmospheric conditions can affect data quality and must therefore be taken into account during analysis. The \ha data consist of images with a resolution of 2048$\times$2048 pixels and a nominal spatial resolution of approximately 2 arcseconds. Each telescope produces one image per minute. The network achieves an overall duty cycle of about 90\%; however, interruptions caused by instrumental issues or unfavorable sky conditions occasionally lead to missing images, resulting in data gaps within the time series.

\subsection{Data selection}\label{sec:data-selection}
We analyze temporal sequences of GONG \ha images, organizing the data into full-day bins. Although this binning scheme is somewhat arbitrary, it facilitates the grouping of detections on a daily basis. The selected time sequence duration also defines the upper limit of detectable oscillation periods. Nevertheless, most filament oscillations exhibit periods shorter than a few hours \citep[see, e.g.,][]{arregui_prominence_2018, luna_gong_2018}. In future analyses, longer temporal bins spanning several days will be employed to investigate oscillations with extended periods, such as those reported by \citet{foullon_detection_2004, foullon_ultra-long-period_2009} and \citet{efremov_ultra_2016}.

We constructed a continuous \ha time series by merging data from the different telescopes in the GONG network. We developed an algorithm to select, at each minute interval, the telescope that provides the best image quality. We focused on ensuring minimal switching between instruments and maximizing instrument continuity to reduce corrections due to telescope intensity level differences and misalignment between the instruments.
More details about the specific algorithm we used can be seen in \citet{castello2025}. 

\subsection{Data preprocessing}\label{sec:data-preprocessing}
Once the available images for a day of observations have been selected and assembled into a single sequence, we start the preprocessing pipeline to prepare it for further analysis.
We corrected each \ha image for solar limb darkening by deriving a robust radial intensity profile from median values in concentric bins across the solar disk. This profile was fitted with a low-order polynomial to obtain a smooth, azimuthally symmetric limb-darkening model.
Following limb-darkening correction, we remove any residual large-scale intensity gradients by fitting a smooth 2D surface to the solar disk. This background model is constructed using Zernike polynomials (with $|m| \le 2$) in normalized polar coordinates. To ensure the model represents only the background and is not biased by resolved solar structures, the fit is performed in log-intensity space using iterative sigma-clipped least squares on a subsampled set of pixels. Finally, we flatten the image by dividing the original frame by the evaluated Zernike model and renormalize it so that the median on-disk intensity is approximately unity.
Next, images are aligned to a 12:00~UT reference frame using a Python-adapted version of the SolarSoft WCS mapping algorithms\footnote{ See \url{https://github.com/dgary50/mapping} for the specific implementation.}. Finally, all off-disk pixels are masked, effectively excluding prominences and restricting the analysis exclusively to on-disk filaments.
A qualitative visualization of these preprocessing stages is provided in Appendix~\ref{appendix:preprocessing-visualization}, where representative frames illustrate the effect of the initial selection, correction, outlier rejection, alignment, and masking steps. This overview complements the methodological description by showing how frames of different initial quality are transformed or discarded by the pipeline.

\subsection{Filament detection and segmentation}
Automatic detection and segmentation of solar filaments in \ha\ images remains challenging because filaments exhibit irregular morphology, variable contrast across the solar disk, and sensitivity to noise and observational artifacts. Early studies in the 2000s relied mainly on classical image-processing techniques, including global thresholding, region growing, and morphological operations, to isolate filament structures \citep{gao2002development, shih2003automatic, fuller2004automatic, bernasconi2005advanced}. These methods achieved useful results, but they typically required manual parameter tuning and often struggled with fragmented or low-contrast filaments. Later work introduced adaptive thresholding and statistical preprocessing to better account for center-to-limb variations and background fluctuations \citep{atoum2013automated, hao2013developing}. The transition to machine learning began with classical models such as artificial neural networks \citep{zharkova2005filament} and support vector machines \citep{qu2005automatic}, which showed that filament identification could benefit from data-driven feature extraction. In the last decade, deep-learning approaches have substantially advanced the field, with convolutional neural networks, and especially U-Net-based architectures, achieving strong performance in the segmentation of complex filament structures \citep{zhu2019solar, liu2021solar}. More recent developments include attention-based models \citep{liu2021solar}, instance-segmentation frameworks \citep{guo2022solar}, and semi-supervised pipelines designed to improve robustness and generalization across datasets \citep{diercke2024universal}. In particular, \citet{reche2024tracking} combined a DETR-based model for filament detection and classification, a U-Net for instance segmentation, and a dedicated tracking algorithm, reporting state-of-the-art performance on their benchmark across these tasks. Overall, these advances have made large-scale, automated processing of solar observations increasingly feasible, with direct applications to statistical studies of solar activity and space-weather monitoring.

In this work, we use the detection and segmentation models from \citet{reche2024tracking} due to their strong performance in both tasks.
The detection model provides bounding boxes for the filaments present in each frame. Within each bounding box, the segmentation model produces a binary mask that separates filament pixels from background pixels.

\begin{figure}[!ht]
    \centering \includegraphics[width=0.43\textwidth]{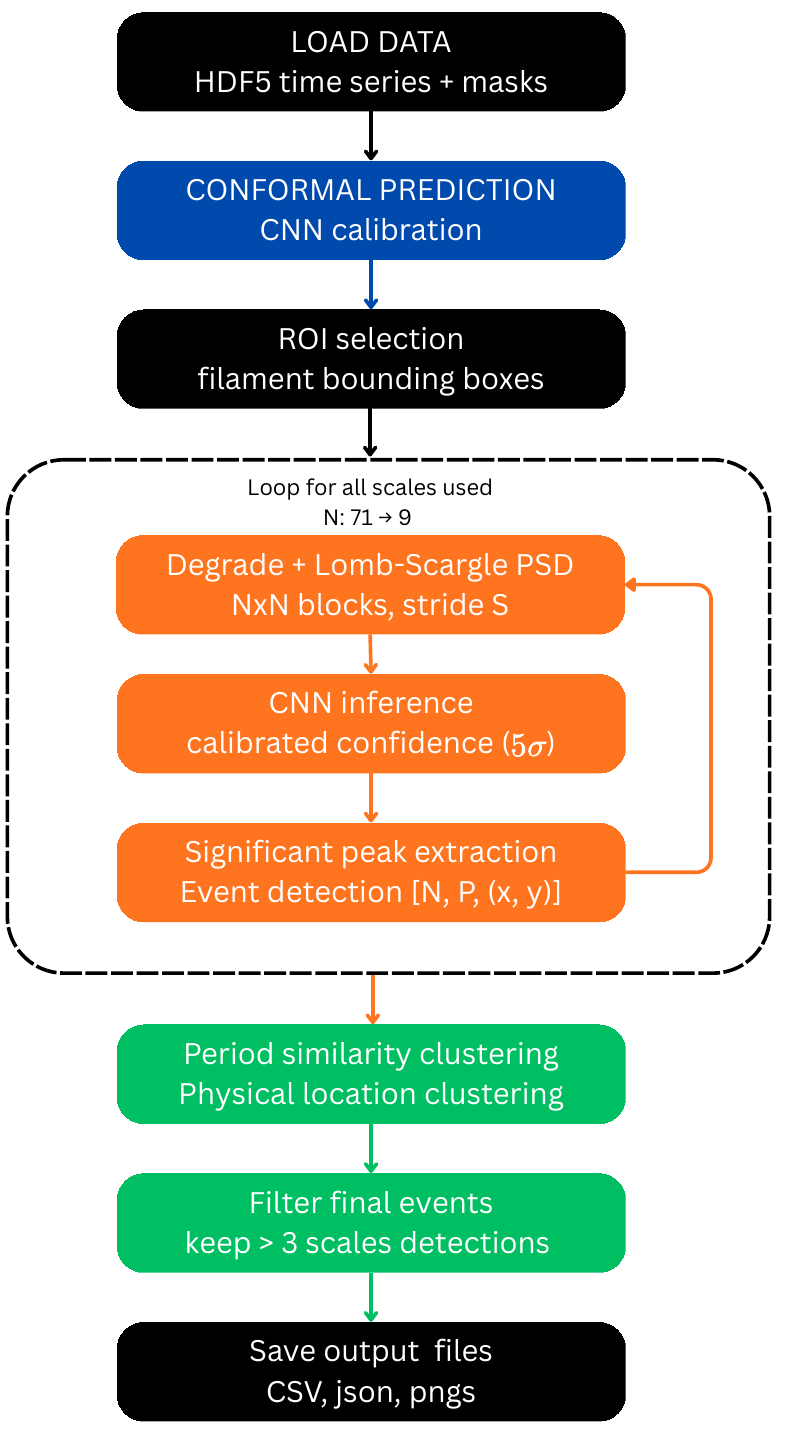}
    \caption{Schematic overview of the oscillation-detection pipeline.}
    \label{fig:pipeline}
\end{figure}

\section{Automatic oscillation analysis pipeline}\label{pipeline}
In this section, we describe the pipeline used for the automatic detection of oscillatory motions in solar filaments.
Our method builds on the framework introduced by \citet{castello2025}, but introduces two main modifications. First, the fixed-confidence threshold used in that work is replaced by a calibrated, pixel-dependent significance threshold. Second, detections are consolidated through a multi-scale procedure designed to recover spatially coherent oscillatory events.
A summarized representation of the developed pipeline can be seen in \cref{fig:pipeline}.

\subsection{Conformal calibration}
In \citet{castello2025}, the neural network was trained to predict a significance threshold at a fixed confidence level. While effective, such a formulation ties the operating point to the training stage and does not clearly separate the estimation of the smooth spectral background from the calibration of statistical significance.
Here, we adopt a conformal prediction (CP)-inspired calibration strategy. Rather than using conformal prediction to construct prediction intervals directly, we use the same principle to calibrate the residuals between the observed power spectra and the CNN-predicted background model. This yields an empirical, day-dependent correction that transforms the model background into a significance threshold. A more detailed description is provided in Appendix \ref{appendix:cp}. As usual, strict conformal guarantees would require exchangeability assumptions, so the method should be interpreted here as an empirical calibration procedure inspired by conformal prediction.

\subsection{Detection pipeline}
For each observing day, the input consists of a time-ordered image cube,
\begin{equation}
I(t_k,x,y), \qquad k=1,\dots,T,
\end{equation}
and a corresponding binary mask cube,
\begin{equation}
M(t_k,x,y)\in\{0,1\},
\end{equation}
identifying filament pixels. The calibration step is performed once per day and subsequently applied to all filament candidates detected in that dataset.

Candidate filaments are defined as connected components in the first mask frame. Only regions larger than 750 pixels are retained. For each selected component, a padded bounding box defines a region of interest (ROI), within which all subsequent analysis is carried out. A persistent support mask is then constructed from the full mask sequence by retaining pixels that remain active for at least 360 frames and applying a small dilation. This support is used to restrict the analysis to locations that remain consistently associated with the filament body.

Each ROI is analyzed at multiple spatial scales. For a given scale \(N\), the image sequence is spatially degraded by averaging over overlapping \(N\times N\) patches, with stride
\begin{equation}
S = \max\!\left(1,\ \mathrm{round}\bigl(N(1-\rho)\bigr)\right),
\end{equation}
where the overlap fraction is fixed to \(\rho=0.75\). The range of analyzed scales extends from \(N_{\max}=71\) to \(N_{\min}=\) 9
, retaining only scales that produce distinct degraded grid dimensions. At each scale, the degraded time series are mean-centered, and the persistent support mask is degraded in the same way to estimate the fraction of filament coverage in each degraded pixel. Only pixels with at least \(30\%\) overlap with the support mask are retained for spectral analysis.

For each valid degraded pixel, we compute a Lomb--Scargle power spectrum on a fixed frequency grid. The use of Lomb--Scargle is motivated by its robustness to the non-uniform sampling that may arise in ground-based observations. The smooth spectral background is modeled as
\begin{equation}\label{eq:noise_model}
\mathcal{S}(f;a,\alpha,b)=a\,f^{-\alpha}+b,
\end{equation}
where \(a\) is the amplitude of the red-noise component, \(\alpha\) its slope, and \(b\) an additive white-noise floor. These parameters are inferred from each spectrum using the pre-trained one-dimensional CNN of \citet{castello2025}.

The statistical threshold is obtained by combining the CNN background estimate with a day-level empirical correction derived from calibration pixels outside filament regions. For a calibration spectrum \(\mathcal{P}_i(f)\), the CNN provides a fitted background \(\widehat{\mathcal{S}}_i(f)\), from which we compute log-residuals
\begin{equation}
\mathcal{R}_i(f)=\log \mathcal{P}_i(f)-\log \widehat{\mathcal{S}}_i(f).
\end{equation}
These residuals are summarized frequency by frequency using robust location and scale estimates, standardized, and converted into an upper empirical quantile. In the implementation used here, the tail probability is fixed to $\delta = 6\times10^{-7}$, approximately corresponding to a two-sided \(5\sigma\) Gaussian tail. This defines a day-dependent multiplicative correction \(\mathcal{C}(f)\), and the final significance threshold for degraded pixel \((i,j)\) is written as
\begin{equation}
\tau_{ij}(f)=\widehat{\mathcal{S}}_{ij}(f)\,\mathcal{C}(f).
\end{equation}
The threshold is therefore pixel-dependent through the CNN background estimate, but calibrated at the day level through the empirical residual distribution. Peak detection is restricted to the interval $f\in\left[9.26\times10^{-5},\,10^{-3}\right]\ \mathrm{Hz}$, corresponding to periods between 3 h and 16.67 min.
At each spatial scale, all significant periods are grouped using a hybrid absolute-plus-relative tolerance criterion: two periods are considered compatible if they differ by no more than 5 min in absolute value or \(5\%\) in relative value. This yields a set of scale-dependent period groups. For each group, the corresponding degraded pixels are projected back onto the degraded spatial grid and merged using connected-component labeling, so that each connected component represents a spatially coherent oscillatory patch at that scale.

After all scales have been processed, the resulting components are pooled and clustered again in period, using the same tolerance rule, to define global period families. These family group detections recur at similar periods across scales, but they do not necessarily correspond to individual physical events. To recover candidate oscillation events, components within the same family are linked according to spatial consistency. Two detections are connected when either their full-image bounding boxes have an Intersection-over-Union metric (IoU) above a defined threshold of $\mathrm{IoU} \geq 0.05$, or their full-image centroids are close enough depending on the scale of each centroid.

Each event is summarized by its representative period and spatial extent, together with the number of contributing detections and distinct spatial scales. In the final catalog, we retain only events supported by at least four different scales.

\section{Results and comparison with Luna et al. (2018)}
\begin{figure}[!ht]
    \centering
    \includegraphics[width=0.45\textwidth]{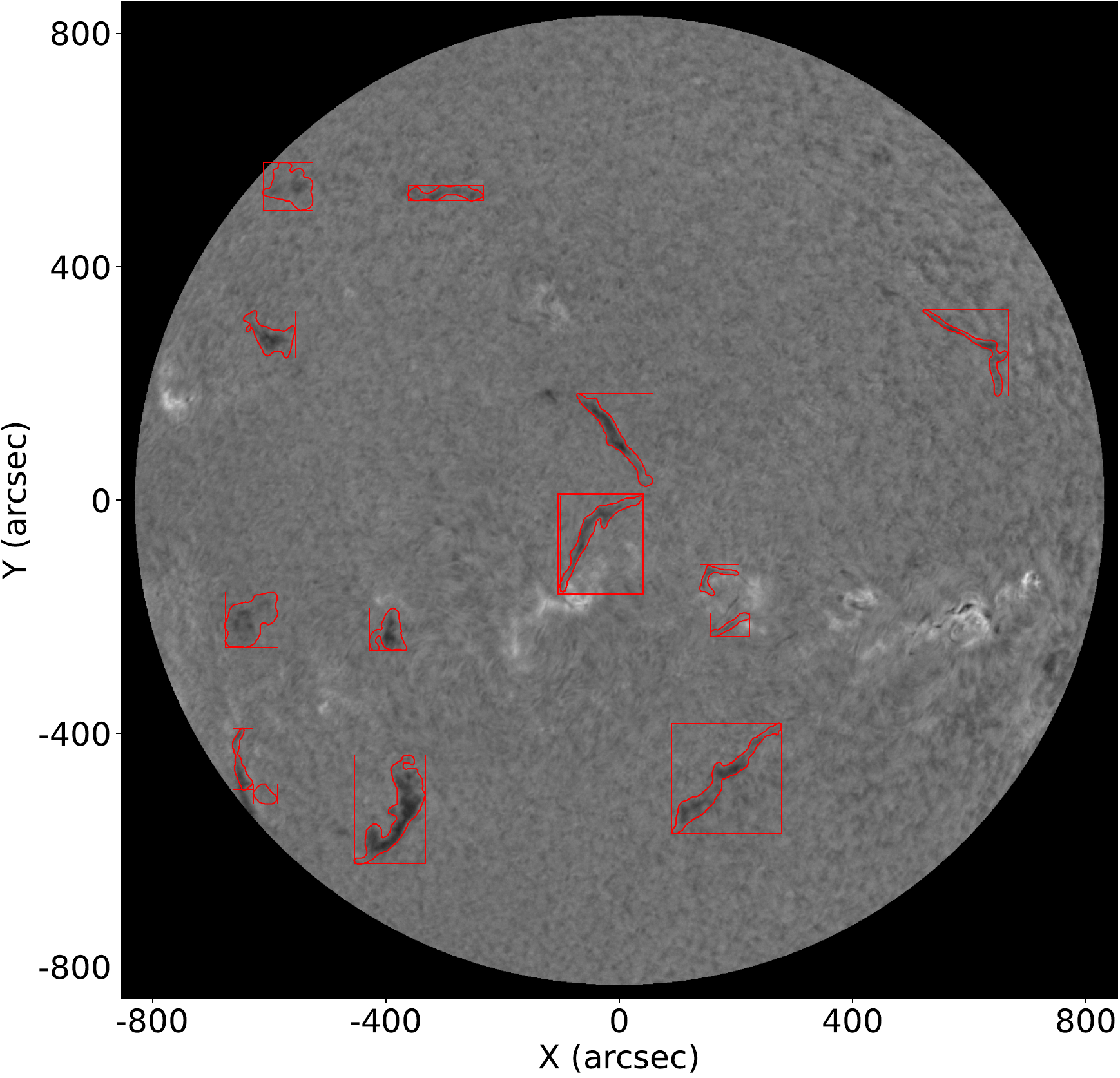}    
    \centering\includegraphics[width=0.45\textwidth]{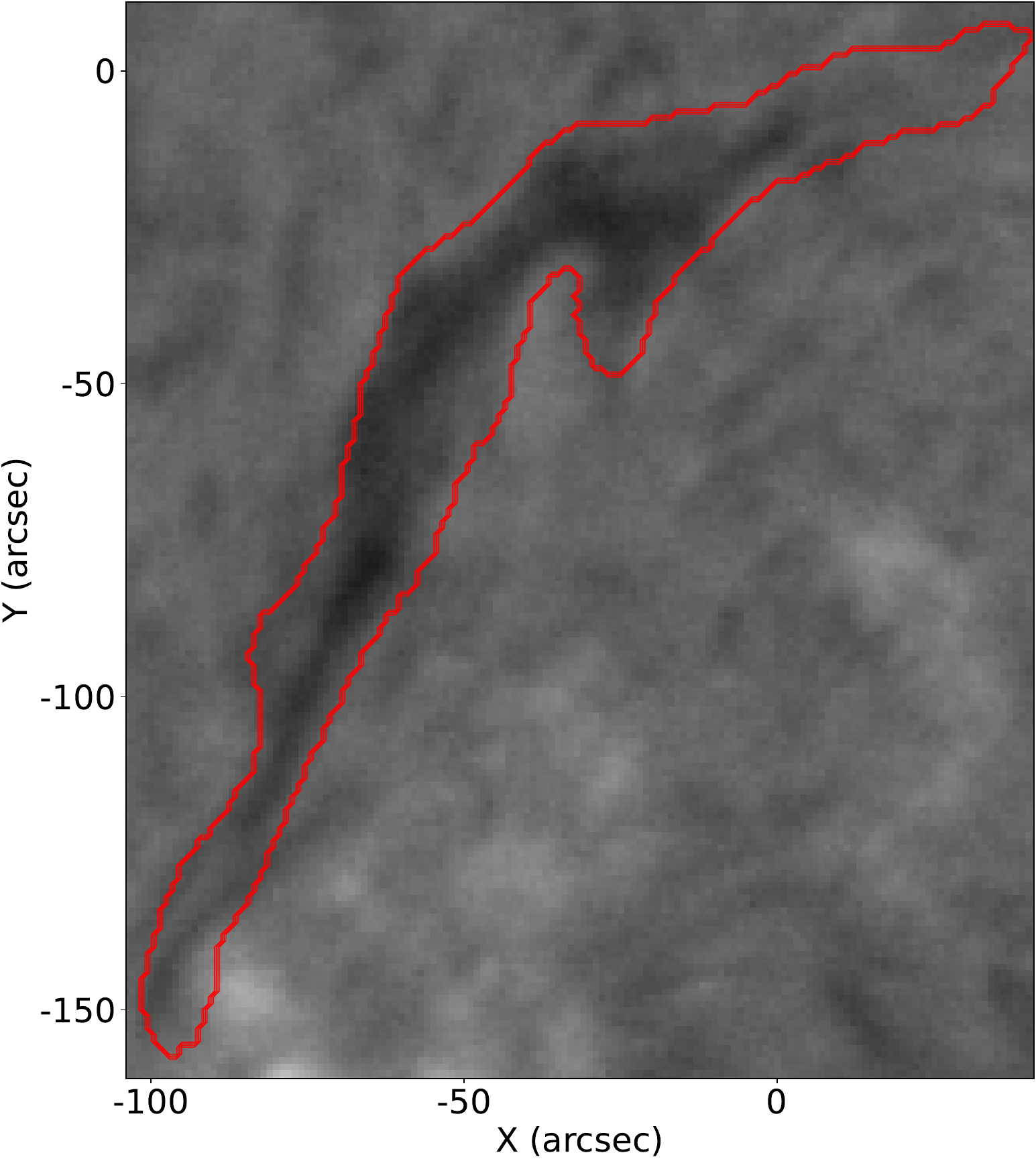}
    \caption{Processed frame from January 1st 2014 around 12:00 UTC. Left: Full disk image from our data preprocessing pipeline, with the corresponding bounding boxes and boundaries from the segmentation model. Right: Zoomed-in view of the filament of interest for the first event, with the corresponding segmentation boundary in red for clarity. It corresponds to the filament located at the solar disk center with a thick bounding box.}
    \label{fig:20140101_zoomed}
\end{figure}

In this section, we show a detailed example of our pipeline results and a comparison of our findings with the results presented in \citet{luna_gong_2018}. Specifically, we compare the results for the first two weeks of January 2014.

\subsection{January 1st, 2014}\label{20140101_detail}
We present a detailed analysis of an oscillatory event detected by our pipeline, comparing it directly with the results reported by \citet{luna_gong_2018}.

This case, which occurred on January 1st 2014, involves a filament located near the disk center with its southern footpoint rooted close to an active region (\cref{fig:20140101_zoomed}). The event corresponds to the first entry in the catalog of \citet{luna_gong_2018}, where the authors employed manually placed slits to construct the time--distance diagrams used for their analysis. In their original work, the filament and the slit geometry are shown in their Fig.~3, while the corresponding time--distance diagram is presented in their Fig.~4a. The oscillation was associated with a nearby flare at approximately 13:50~UT and was reported to have a period of $75.9 \pm 1$ minutes.

The pipeline first produces a set of spatially degraded versions of the image sequence by applying square averaging kernels of size $N$ with stride $S$. Each degraded pixel therefore represents the mean intensity within a local region of the filament. For every degraded pixel, we compute the PSD of the corresponding intensity time series and compare it with the CNN-predicted background model introduced by \citet{castello2025}. Significant spectral peaks are then identified using the CP-based daily thresholds described in Appendix \ref{appendix:cp}.

Figure \ref{fig:collage_different_N} illustrates this procedure for three representative spatial scales, $N=44$, $35$, and $16$, in the region where the oscillation reported by \citet{luna_gong_2018} was originally analyzed with a manually placed slit. The left column shows the averaging kernel, shown as a red grid, centered on the same approximate filament location. The middle column shows the resulting averaged intensity time series, and the right column shows the corresponding PSD. In each PSD, the red dashed curve denotes the CNN-predicted background level, while the blue dashed curve denotes the CP-derived significance threshold for that observing day.
\begin{figure*}[!ht]
    \centering
    \includegraphics[width=\textwidth]{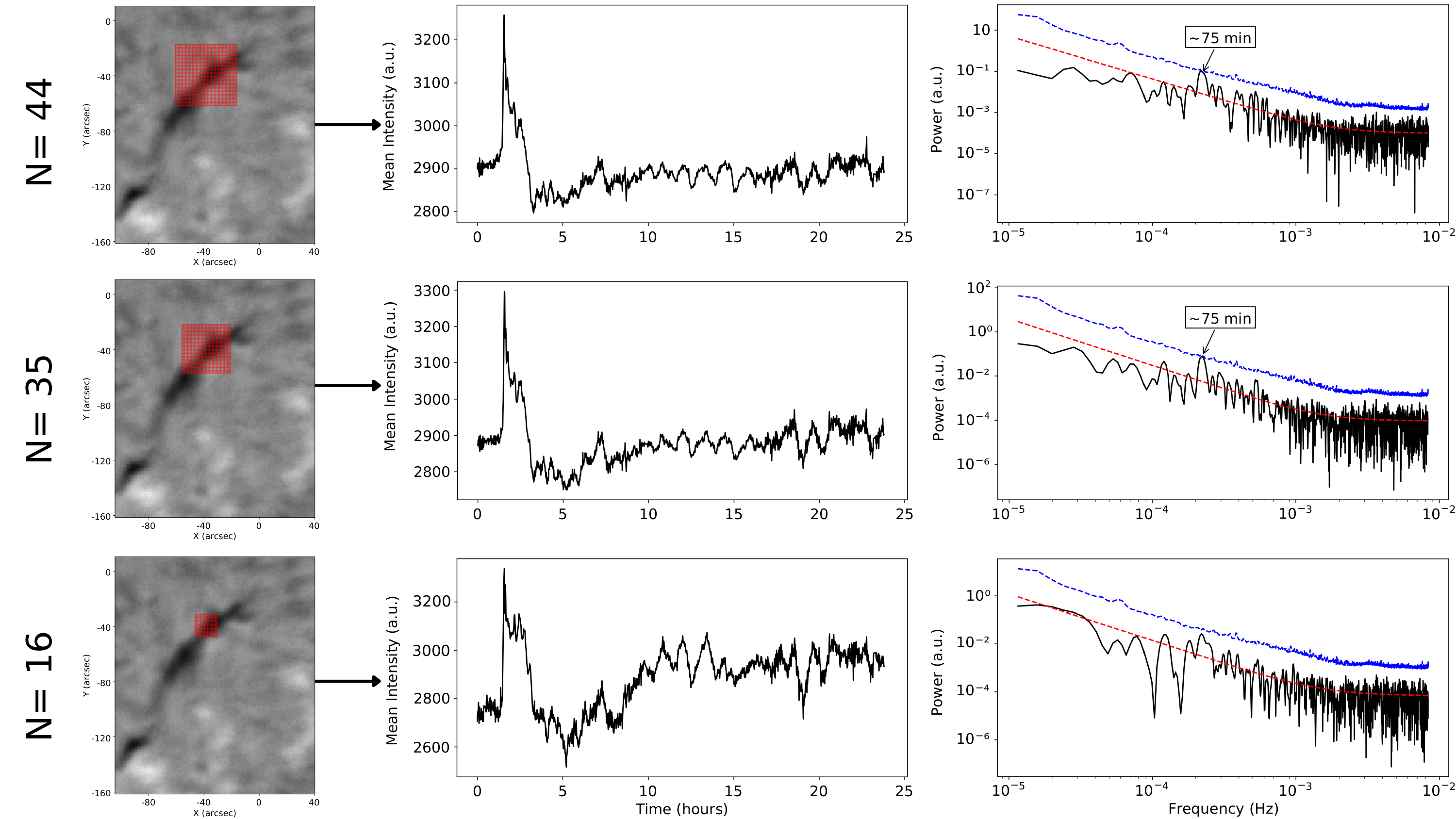}
    \caption{Visual representation of the multi-scale processing applied to the filament shown in Fig. \ref{fig:20140101_zoomed}. Left column: position of the spatial averaging kernel within the ROI. Middle column: averaged intensity time series obtained from that kernel. Right column: corresponding PSD, together with the CNN-predicted background model (red dashed line) and the CP-derived significance threshold (blue dashed line). From top to bottom, the rows correspond to kernel sizes $N=44$, $35$, and $16$, respectively.}
    \label{fig:collage_different_N}
\end{figure*}

A clear scale dependence is already visible in Fig. \ref{fig:collage_different_N}. The large and intermediate kernels show a significant peak at a period of approximately $75$~min, whereas the smallest kernel, $N=16$, does not yield a significant detection. This behavior can be interpreted in terms of spatial coherence. When the averaging kernel covers a sufficiently large fraction of the coherently oscillating filament region, small-scale intensity fluctuations are partially averaged out and the dominant collective displacement becomes more visible in the PSD. Conversely, at finer spatial scales, the degraded time series is more strongly affected by local filament substructure and by small-scale motions that are not necessarily phase-coherent with the global oscillation. These local contributions can dominate the intensity evolution and reduce the height of the spectral peak associated with the large-scale mode.

This effect is directly related to one of the main limitations identified in \citet{castello2025}: the high spatial resolution of the data contains a large amount of fine-scale variability, making it difficult to distinguish global filament oscillations from local motions. The present multi-scale approach addresses this limitation by explicitly searching for signals that remain spatially coherent over a range of averaging scales.

To quantify the scale dependence suggested by Fig. \ref{fig:collage_different_N}, we performed an a posteriori SNR analysis of the 1 January 2014 event. 
Upon processing this region, the pipeline successfully identifies a spatially coherent oscillation and defines an event bounding box that encapsulates the pixels where the oscillation signal is detected across the different scales.
Using the geometric center of this bounding box as a reference point, we recomputed the degraded intensity time series, its PSD, and the corresponding CNN-predicted background model for each kernel size $N$.
We then evaluated the spectral SNR at the detected oscillation period, $P_0 = 75.9$~min, as
\begin{equation}
    \mathrm{SNR}(N) = 
    \frac{\mathcal{P}_N(f_0)}{\hat{\mathcal{S}}_N(f_0)},
    \qquad f_0 = \frac{1}{P_0},
\end{equation}
where $\mathcal{P}_N(f_0)$ is the PSD value at the oscillation frequency and $\hat{\mathcal{S}}_N(f_0)$ is the CNN-predicted background level at the same frequency.

To ensure this reference point provides a representative measure of the oscillation, we also assessed the sensitivity of the SNR to the exact kernel placement. The calculation was repeated across a small grid by shifting the kernel center within a $\pm 5$~pixel neighborhood in both spatial directions. This produces, for each scale~$N$, a distribution of SNR values that reflects the impact of spatial sampling and centering uncertainties within the detected oscillatory region.
\begin{figure}[!ht]
    \centering
    \includegraphics[width=1\linewidth]{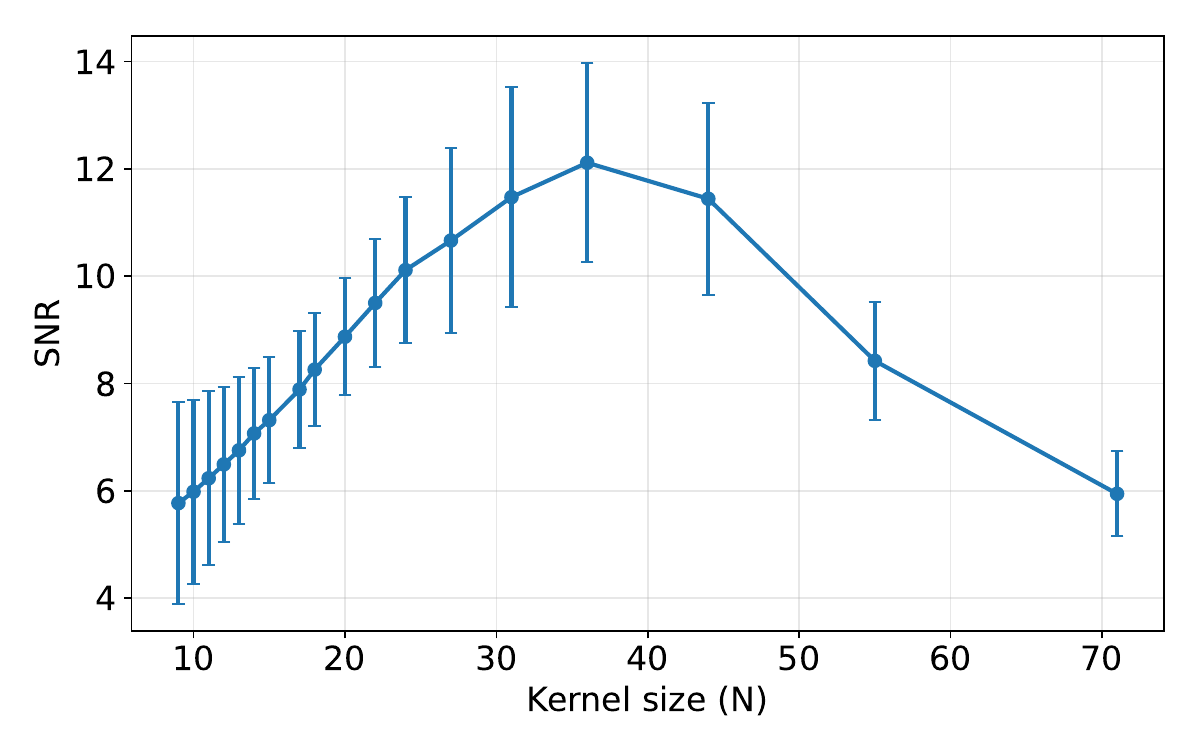}
    \caption{Spectral SNR of the 1~January 2014 oscillation as a function of the averaging kernel size~$N$. The SNR is calculated at the detected period, $P_0 = 75.9$~min, as the ratio between the PSD power and the CNN-predicted background level. Points represent the mean SNR obtained by repeating the analysis for shifted kernel centers around the geometric center of the detected event. Error bars indicate the standard deviation of the SNR distribution at each scale.}
    \label{fig:SNR_vs_N}
\end{figure}

\begin{figure*}[!ht]
    \centering
    \includegraphics[width=\linewidth]{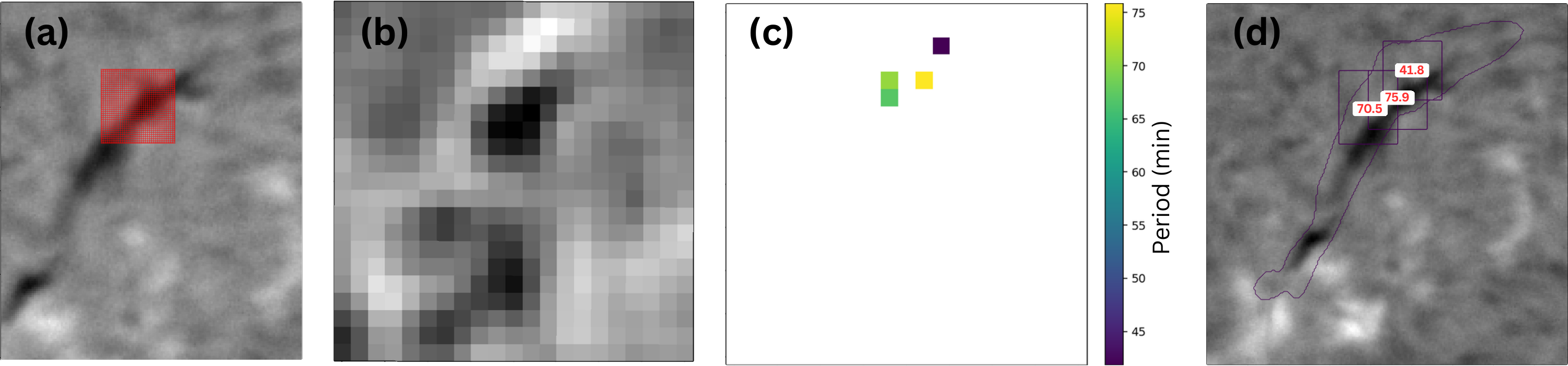}
    \caption{Visual representation of our pipeline focusing on a single scale ($N=35$, middle row in \cref{fig:collage_different_N}). (a): \ha intensity image with the used averaging kernel represented in red. (b): Degraded image produced using the shown kernel in (a). (c): Significant periods found in the degraded image sequence. Only four degraded pixels have significant periods. (d): Representation on the original image, of the bounding boxes of the significant periods found in the degraded sequence, with the corresponding period of each bounding box in the center.}
    \label{fig:one_scale_example}
\end{figure*}

Figure~\ref{fig:SNR_vs_N} shows the mean SNR as a function of kernel size, with error bars representing the standard deviation across the various kernel placements. The curve displays a characteristic non-monotonic dependence on the spatial averaging scale: at small scales, the SNR is relatively low, indicating that the intensity time series is dominated by local fine-scale dynamics and incoherent fluctuations. As~$N$ increases, these localized fluctuations are progressively suppressed, allowing the spectral signature of the oscillation to become more prominent until reaching a peak at intermediate kernel sizes ($N = 35$). Beyond this optimal range, the SNR decreases as the larger averaging aperture begins to incorporate non-oscillating filament material and surrounding background regions. This inclusion dilutes the oscillatory signal and reduces the contrast of the spectral peak, as these regions do not share the same coherent motion. Consequently, this behavior provides a quantitative justification for our multi-scale strategy, demonstrating that the detectability of an oscillation is maximized when the averaging kernel size is comparable to the intrinsic spatial scale over which the filament motion remains coherent.

While \cref{fig:collage_different_N} displays a single pixel of interest, the full analysis is performed systematically for all pixels across every scale. Figure \ref{fig:one_scale_example} illustrates the pipeline output specifically for the $N=35$ scale. 
The sequence from left to right illustrates the data progression through the pipeline. 
The kernels defined by $N$ and $S$ are applied across the entire image sequence to generate the degraded stack shown in \cref{fig:one_scale_example}a and b. 
Spectral analysis is then performed on each degraded pixel, recording its spatial coordinates alongside the identified period (\cref{fig:one_scale_example}c).
In this example, four degraded pixels yield significant detections in the northern region of the image, with periods of 41.8, 69.8, 71.2, and 75.9 minutes.
Finally, panel (d) illustrates the candidate event bounding boxes for this scale, mapped back onto the original image sequence.
Note that the 69.8 and 71.2-minute periods are close in both period and spatial location; consequently, they are fused into a single bounding box with a mean period of 70.5 minutes.

Following this, we apply a cross-scale consistency criterion to ensure the robustness of the identified oscillations. Only candidate events that coexist across at least four different scales in both spatial location and period are classified as confirmed detections and included in the final results. This multi-scale persistence requirement filters out localized transients and ensures that only spatially and spectrally coherent events are recovered.

\subsection{Validation against the \citet{luna_gong_2018} Catalog}

\begin{table}[!ht]
\caption{\label{tab:luna_catalog}Summarized first two weeks of events from the catalog presented in \citet{luna_gong_2018}. EN refers to the catalog event number, NFF means that the filament was not detected in the first frame of the time series, NOF signifies that no oscillation was found using our pipeline, NES means that an event was found similar to the one reported, but it is not supported by enough scales and finally DPF means that different periods were found using the new pipeline. Bold entries highlight events reported in both catalogs.}
\centering
\small
\setlength{\tabcolsep}{1pt}
\begin{tabular}{c c c c c c c}
\hline\hline
EN & Date & Time & $x$ (arcsec) & $y$ (arcsec) & Period (min) & Disagreement \\
\hline
\textbf{1} & \textbf{2014-01-01} & \textbf{13:50} & \textbf{-17.8} & \textbf{-75.9} & $\textbf{75.7} \pmb{\pm} \textbf{1.0}$ & \textbf{-} \\
3 & 2014-01-04 & 20:30 & -251.4 &  301.7 & $57.0 \pm 1.1$ & NFF \\
4 & 2014-01-05 & 11:43 & -398.6 &   48.4 & $63.2 \pm 3.4$ & NOF\\
\textbf{5} & \textbf{2014-01-05} & \textbf{12:01} & \textbf{-308.3} & \textbf{-115.6} & $\textbf{52.1} \pmb{\pm} \textbf{0.8}$ & \textbf{-} \\
7 & 2014-01-05 & 10:44 &  730.0 & -430.2 & $63.2 \pm 2.6$ & NFF\\
\textbf{9} & \textbf{2014-01-05} & \textbf{20:13} &  \textbf{805.9} &  \textbf{-78.6} & $\textbf{44.1} \pmb{\pm} \textbf{2.0}$ & \textbf{-}\\
10 & 2014-01-06 & 06:58 &   84.4 &  491.8 & $68.5 \pm 0.5$ & NFF \\
11 & 2014-01-06 & 08:02 & -132.3 & -261.2 & $62.7 \pm 1.6$ & NES \\
12 & 2014-01-06 & 11:07 & -267.8 & -466.3 & $39.9 \pm 2.2$ & NOF\\
13 & 2014-01-06 & 12:27 & -184.5 &   57.4 & $60.5 \pm 0.6$ & NOF\\
15 & 2014-01-06 & 06:45 &   57.0 &  303.9 & $63.8 \pm 0.5$ & NFF\\
\textbf{16} & \textbf{2014-01-07} & \textbf{05:58} & \textbf{-334.6} & \textbf{78.3} & $\textbf{64.7} \pmb{\pm} \textbf{1.3}$ & \textbf{-}\\
17 & 2014-01-07 & 04:04 & -324.9 & -282.1 & $46.2 \pm 0.9$ & NOF\\
\textbf{18} & \textbf{2014-01-07} & \textbf{04:45} &  \textbf{142.6} & \textbf{-369.6} & $\textbf{49.9} \pmb{\pm} \textbf{0.9}$ & \textbf{-}\\
19 & 2014-01-07 & 14:16 &  -99.6 & -501.0 & $43.5 \pm 0.3$ & NES\\
20 & 2014-01-08 & 05:47 & -139.5 & -316.0 & $49.6 \pm 1.1$ & DPF\\
21 & 2014-01-08 & 04:49 & 7.9 & -494.3 & $42.9 \pm 1.3$ & DPF\\
23 & 2014-01-08 & 03:28 &  259.6 & -367.1 & $58.6 \pm 0.7$ & NOF\\
24 & 2014-01-09 & 16:56 &  155.4 & -305.8 & $51.3 \pm 1.2$ & NOF\\
25 & 2014-01-09 & 12:31 &  241.8 & -504.8 & $46.5 \pm 2.6$ & NOF\\
26 & 2014-01-10 & 11:23 &  432.5 &   80.5 & $42.8 \pm 1.1$ & DPF\\
27 & 2014-01-11 & 12:37 &  815.4 & -392.9 & $65.5 \pm 2.9$ & NOF\\
\hline
\end{tabular}
\end{table}

We applied our pipeline to the first two weeks of GONG H$\alpha$ observations from January 2014. This interval allows for a direct comparison with the events reported in the \citet{luna_gong_2018} catalog. This focused timeframe serves to demonstrate the detection capabilities and the reliability of our methodology. While this work focuses on the validation of the pipeline, a comprehensive statistical analysis of the full GONG dataset will be presented in a subsequent study.

Table \ref{tab:luna_catalog} summarizes the events reported by \citet{luna_gong_2018} during the first two weeks of 2014, where bold entries denote those successfully recovered by our pipeline.
This table lists the event index, observation time, filament coordinates (arcsec), and the period reported by \citet{luna_gong_2018}, followed by a final column detailing any discrepancies identified in our comparison. 
We identify four primary sources of discrepancy between the two studies. First, No Filament Found (NFF) refers to cases where the filament is not identified by our detection model in the initial frame of the image sequence.
Second, No Oscillation Found (NOF) occurs when the pipeline detects no significant oscillation in the analyzed time series, an expected outcome given the fundamental differences between the two methodologies. Third, Not Enough Scales (NES) applies to detections that match the reported spatial and spectral parameters but fail to meet our cross-scale consistency criterion. Finally, Different Period Found (DPF) denotes instances where the pipeline identifies a global oscillation mode with a period differing from the one previously reported.
No minimum period-separation threshold was imposed for DPF cases. Instead, the previously reported cases were manually and empirically cross-checked against the newer catalog. All cases categorized as the same event in both catalogs have period differences below 5 minutes. By contrast, the cases classified as DPF were unambiguous, with discrepancies on the order of tens of minutes, making a specific numerical threshold unnecessary for this validation.

Table \ref{tab:luna_catalog} shows that our pipeline successfully recovered six of the events reported by \citet{luna_gong_2018}. The remaining discrepancies are categorized into four primary sources. First, NFF (4 cases) highlights the inherent challenges of automated segmentation in initial frames. Second, the NOF instances (8 cases) suggest that some oscillations identified through manual slit placement may not meet our pipeline's stricter spectral significance criteria. Furthermore, the low incidence of NES (2 cases) and DPF (3 cases) indicates that when a signal is captured, it remains both statistically robust across scales and consistent in period with previously reported values. These discrepancies do not imply a lack of reliability; rather, they reflect the transition from subjective manual inspection to a more objective and rigorous automated framework.

\begin{figure}
    \centering
    \includegraphics[width=\linewidth]{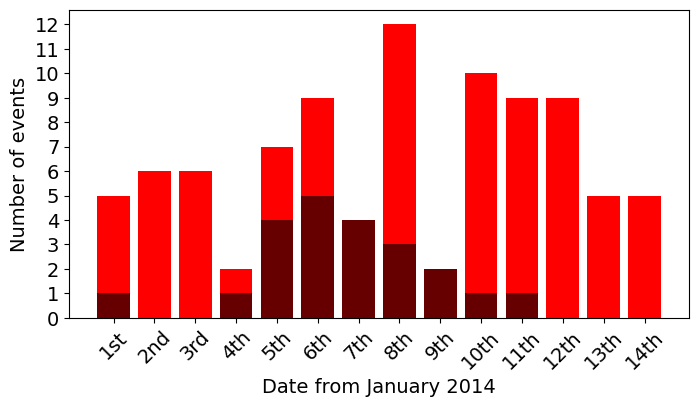}
    \caption{Daily distribution of detected filament oscillations during the first two weeks of January 2014. The total number of events recovered by our automated pipeline (red) consistently exceeds those from the manual survey of \citet{luna_gong_2018} (dark red). Notably, our method identifies significant activity on days where no events were previously reported (e.g., Jan 2, 3, 12–14), demonstrating a substantial increase in detection sensitivity and temporal coverage.}
    \label{fig:histogram_per_day}
\end{figure}
Beyond the comparative validation, the full set of oscillatory events identified by our pipeline is presented in Table \ref{tab:events_catalog}. This table details the detection date, spatial coordinates (arcsec), and the image-space bounding box—calculated from a bottom-left (0,0) origin—which delineates the specific filament region where coherent motion was recovered. The oscillation period for each identified event is also provided. The most striking result is the significant discrepancy in the number of detections: our pipeline identifies 91 events during the first two weeks of January 2014, whereas \citet{luna_gong_2018} reports 22 (after removing duplicates). As visualized in \cref{fig:histogram_per_day}, our method achieves a much higher detection rate, averaging 6.5 events per day compared to the 1.5 events found in the manual catalog. 
This more than four-fold increase in the number of identified events indicates that filament oscillations are far more common than previously estimated. These findings suggest that such oscillations represent a near-constant dynamical signature of these structures, which is only fully revealed once the observational biases of manual inspection are removed.

\begin{figure}
    \centering
    \includegraphics[width=0.45\textwidth]{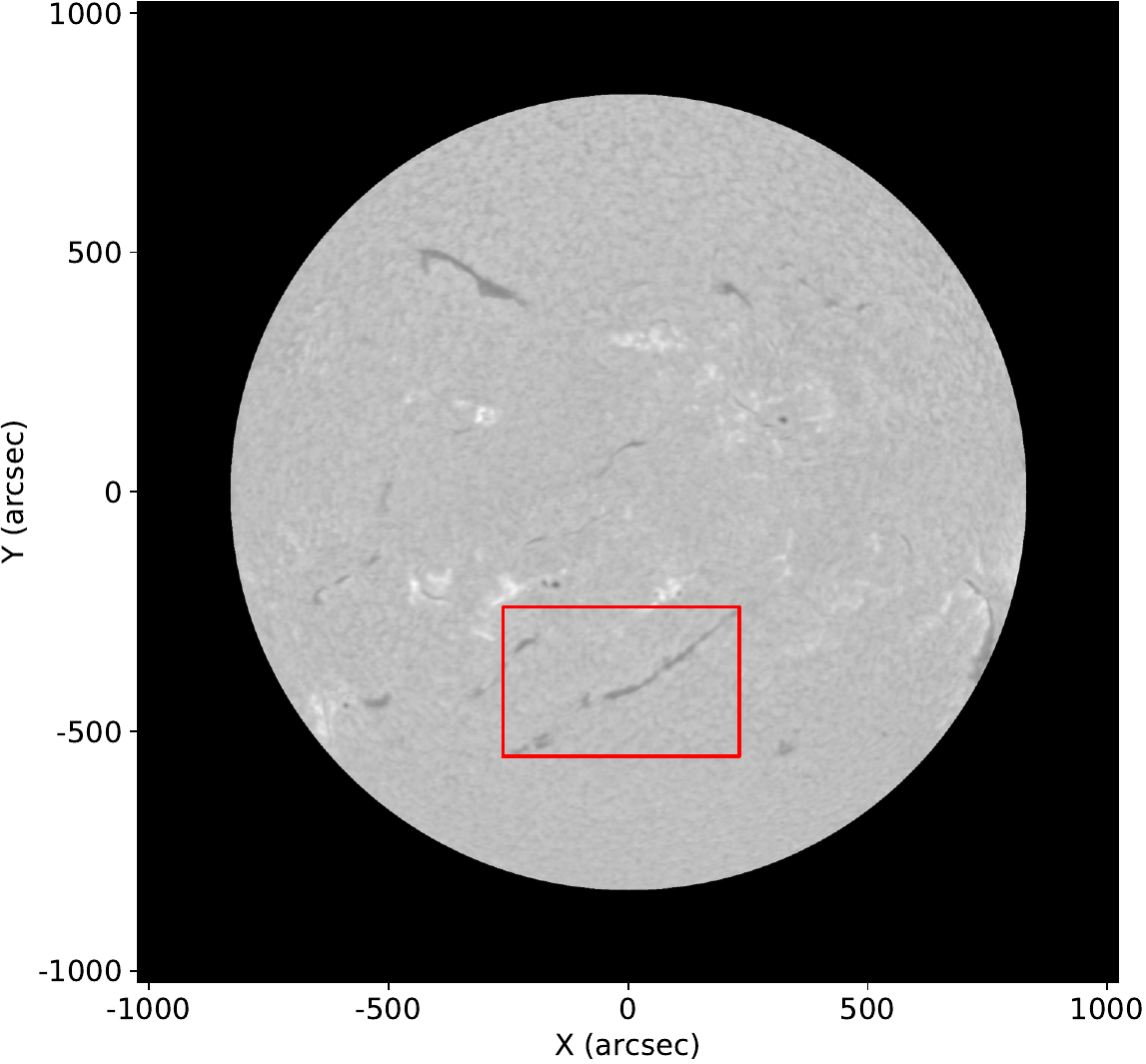}
    \caption{Full-disk \ha image from January 13th 2014. The red box indicates the filament with event number 82 in Table \ref{tab:events_catalog} highlighted in red.}
    \label{fig:20140113_full_disk}
\end{figure}
Notably, the pipeline recovers multiple oscillations on days where \citet{luna_gong_2018} reports no activity. Specifically, we identify six events per day on January 2nd and 3rd, and five each on the 13th and 14th. The most prominent difference occurs on January 8th, where the pipeline resolves twelve distinct oscillatory events compared to only three in the manual record. This demonstrates the pipeline’s ability to reveal a substantial population of events that remained undetected in previous surveys, ensuring complete temporal coverage across the entire two-week period.
\begin{figure*}[!ht]
    \centering
    \includegraphics[width=\linewidth]{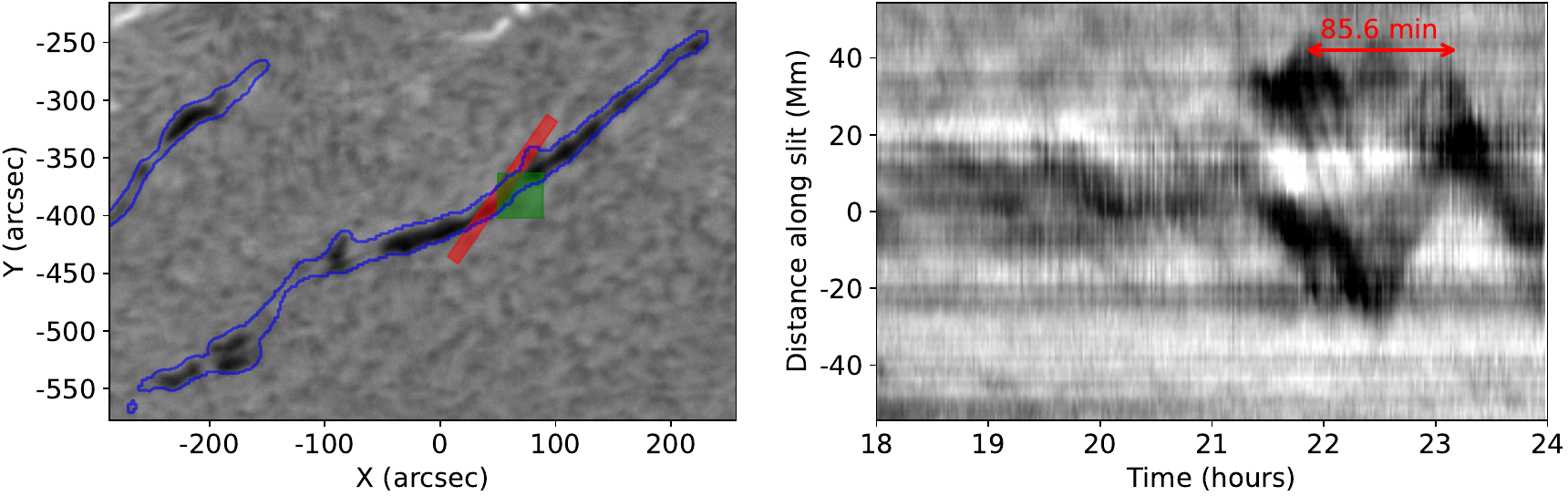}
    \caption{Zoomed in view of EN 82 from Table \ref{tab:events_catalog}. Left: Zoomed-in view of the filament in \ha, the contour of the filament from the segmentation mask (blue), the region where the event is detected (green box), and the slit used to further analyze the event (red rectangle). Right: Time distance diagram produced by the red slit from the left panel. The time distance diagram is zoomed in on the time lapse of the event, and the period is highlighted with a red arrow.}
    \label{fig:20140113_zoom_td}
\end{figure*}
These discrepancies stem from fundamental differences in methodology. The catalog by \citet{luna_gong_2018} relied on the manual inspection of H$\alpha$ movies and the subjective placement of slits where motion was visually apparent. While this traditional approach is a standard in the field, it is naturally biased toward the most conspicuous, large-amplitude oscillations. In contrast, by employing an automated spectral analysis, our pipeline bypasses these observational biases and recovers the underlying bulk of filament dynamics—including low-amplitude or rapid oscillations often overlooked by the human eye—providing a more exhaustive and objective census.

We highlight an event from 13 January 2014 that was absent from the \citet{luna_gong_2018} catalog as a clear example of the pipeline's superior sensitivity. This case involves a large filament in the southern hemisphere, centrally located near the central meridian, see \cref{fig:20140113_full_disk}. In this figure the red bounding box identifies the detected oscillatory region within this structure. By contrasting our automated findings with the traditional manual approach, we demonstrate the system's ability to recover oscillations that are easily overlooked by visual inspection.
This specific detection corresponds to event number (EN) 82 in Table \ref{tab:events_catalog}, which is highlighted in red for clarity. The oscillation is centered at $(x, y) = (67.2, -381.7)$ arcsec and exhibits a period of 85.6 minutes. A detailed view of the filament is provided in the left panel of \cref{fig:20140113_zoom_td}, where the blue contour delineates its segmentation mask. Within the same image, a green box identifies the specific region where the oscillatory signal was recovered by our pipeline.

To validate the detected signal, we performed an independent verification by placing a manual slit across the region identified by the pipeline. This slit, indicated in red in the left panel of \cref{fig:20140113_zoom_td}, was optimized in position and orientation to produce a clear time--distance diagram, displayed in the right panel of the same figure. The resulting diagram reveals an oscillation starting near 22:00~UT with a period of approximately 85 minutes, showing full consistency with our automated results in both spatial location and periodicity. This agreement demonstrates that our pipeline not only recovers events overlooked by manual surveys but also maintains the reliability of traditional methods while providing superior detection sensitivity.
\longtab[1]{
\setlength{\tabcolsep}{5pt}
\begin{longtable}{cccccc}
\caption{Detected oscillation events from our pipeline. The bounding box is reported as $x_{\min}$--$x_{\max}$, $y_{\min}$--$y_{\max}$ in full-frame pixel coordinates. Bold entries highlight cases reported in both catalogs. The entry in red is the example event not reported in Table \ref{tab:luna_catalog}, to validate our pipeline in cases where no direct comparison was possible.}
\label{tab:events_catalog}\\
\hline\hline
EN & Date & $x$ (arcsec) & $y$ (arcsec) & Bounding Box (pix) & Period (min) \\
\hline
\endfirsthead
\caption{Detected oscillation events from our pipeline. The bounding box is reported as $x_{\min}$--$x_{\max}$, $y_{\min}$--$y_{\max}$ in full-frame pixel coordinates. Bold entries highlight cases reported in both catalogs. The entry in red is the example event not reported in Table \ref{tab:luna_catalog}, to validate our pipeline in cases where no direct comparison was plausible.}\\
\hline\hline
EN & Date & $x$ & $y$ & Bounding Box (pix) & Period (min) \\
\hline
\endhead
\hline
\endfoot
\hline
\endlastfoot
1 & 2014-01-01 & -378.7 & -481.4 & 616--668, 523--570 & 75.9 \\
2 & 2014-01-01 & 182.3 & -458.9 & 1197--1216, 557--575 & 58.2 \\
3 & 2014-01-01 & 121.0 & -522.2 & 1138--1152, 496--509 & 58.2 \\
4 & 2014-01-01 & -54.1 & -32.8 & 911--1003, 892--1013 & 68.1 \\
5 & \textbf{2014-01-01} & \textbf{-44.2} & \textbf{-41.3} & \textbf{957--1012, 954--1005} & \textbf{75.9} \\
6 & 2014-01-02 & 213.5 & 114.6 & 1226--1246, 1130--1152 & 63.8 \\
7 & 2014-01-02 & -494.1 & -551.1 & 521--537, 465--480 & 56.6 \\
8 & 2014-01-02 & 129.8 & -107.6 & 1146--1164, 906--927 & 59.1 \\
9 & 2014-01-02 & 218.6 & -16.6 & 1236--1250, 1001--1014 & 59.1 \\
10 & 2014-01-02 & 166.8 & -36.5 & 1179--1205, 974--1002 & 59.1 \\
11 & 2014-01-02 & 213.1 & -8.5 & 1228--1245, 1007--1024 & 50.2 \\
12 & 2014-01-03 & 386.8 & 154.4 & 1404--1418, 1168--1186 & 69.3 \\
13 & 2014-01-03 & 417.4 & 99.7 & 1434--1448, 1118--1131 & 69.3 \\
14 & 2014-01-03 & 447.1 & 78.1 & 1458--1487, 1088--1119 & 78.8 \\
15 & 2014-01-03 & -377.7 & -389.7 & 638--656, 626--644 & 64.8 \\
16 & 2014-01-03 & 367.9 & -17.3 & 1386--1399, 1000--1015 & 62.8 \\
17 & 2014-01-03 & -586.0 & -398.6 & 431--447, 616--632 & 70.5 \\
18 & 2014-01-04 & 526.6 & -80.5 & 1531--1569, 925--962 & 65.9 \\
19 & 2014-01-04 & -612.8 & -13.5 & 405--417, 1004--1017 & 20.2 \\
20 & 2014-01-05 & -306.3 & -376.3 & 710--727, 637--654 & 68.1 \\
21 & 2014-01-05 & -276.0 & -69.2 & 733--763, 945--969 & 37.1 \\
22 & \textbf{2014-01-05} & \textbf{-259.5} & \textbf{-75.4} & \textbf{748--784, 929--970} & \textbf{56.6} \\
23 & 2014-01-05 & -739.2 & -281.3 & 277--293, 732--752 & 46.7 \\
24 & 2014-01-05 & -306.4 & 279.0 & 707--731, 1292--1316 & 45.6 \\
25 & 2014-01-05 & -301.1 & 274.3 & 707--734, 1286--1310 & 67.0 \\
26 & \textbf{2014-01-05} & \textbf{720.7} & \textbf{-38.4} & \textbf{1734--1754, 978--994} & \textbf{44.1} \\
27 & 2014-01-06 & -485.3 & -3.6 & 524--557, 998--1049 & 64.8 \\
28 & 2014-01-06 & -482.1 & -4.6 & 530--556, 1010--1029 & 52.8 \\
29 & 2014-01-06 & -30.6 & 483.0 & 985--1002, 1498--1516 & 60.9 \\
30 & 2014-01-06 & -182.5 & -129.1 & 831--856, 886--913 & 44.1 \\
31 & 2014-01-06 & -58.9 & -67.6 & 955--973, 950--967 & 60.9 \\
32 & 2014-01-06 & -48.5 & -88.3 & 963--988, 926--949 & 53.5 \\
33 & 2014-01-06 & 44.0 & -314.9 & 1061--1076, 703--715 & 50.8 \\
34 & 2014-01-06 & -139.1 & -370.1 & 877--895, 646--666 & 41.4 \\
35 & 2014-01-06 & -137.5 & -369.2 & 881--893, 649--661 & 35.5 \\
36 & \textbf{2014-01-07} & \textbf{-258.5} & \textbf{97.9} & \textbf{747--816, 1102--1160} & \textbf{67.0} \\
37 & 2014-01-07 & -307.6 & 57.1 & 705--728, 1070--1093 & 75.9 \\
38 & \textbf{2014-01-07} & \textbf{121.2} & \textbf{-360.4} & \textbf{1123--1178, 645--686} & \textbf{50.8} \\
39 & 2014-01-07 & 83.6 & -346.9 & 1098--1118, 670--688 & 62.8 \\
40 & 2014-01-08 & -91.7 & -333.2 & 920--944, 677--706 & 40.9 \\
41 & 2014-01-08 & -162.7 & -186.5 & 850--874, 829--849 & 40.9 \\
42 & 2014-01-08 & -73.4 & -287.3 & 928--969, 719--757 & 64.8 \\
43 & 2014-01-08 & -44.6 & 77.3 & 972--987, 1094--1108 & 49.6 \\
44 & 2014-01-08 & 56.8 & -453.9 & 1065--1103, 555--584 & 41.8 \\
45 & 2014-01-08 & 65.2 & -445.7 & 1081--1100, 567--586 & 60.0 \\
46 & 2014-01-08 & 20.9 & -490.1 & 1039--1052, 528--541 & 60.0 \\
47 & 2014-01-08 & 30.3 & -483.0 & 1047--1062, 534--548 & 67.0 \\
48 & 2014-01-08 & 316.4 & 404.5 & 1334--1348, 1421--1436 & 70.5 \\
49 & 2014-01-08 & -503.3 & 374.6 & 499--532, 1387--1418 & 52.1 \\
50 & 2014-01-08 & -492.5 & 376.3 & 513--560, 1387--1413 & 64.8 \\
51 & 2014-01-08 & -44.4 & 76.9 & 972--990, 1093--1108 & 50.2 \\
52 & 2014-01-09 & -321.7 & 370.7 & 696--709, 1387--1401 & 43.6 \\
53 & 2014-01-09 & 135.5 & 46.8 & 1140--1189, 1054--1095 & 63.8 \\
54 & 2014-01-10 & -494.3 & -426.1 & 511--548, 578--614 & 82.1 \\
55 & 2014-01-10 & 309.8 & -241.5 & 1320--1346, 771--799 & 62.8 \\
56 & 2014-01-10 & 308.6 & -268.1 & 1326--1339, 748--762 & 51.5 \\
57 & 2014-01-10 & 79.1 & 467.9 & 1097--1110, 1486--1499 & 43.6 \\
58 & 2014-01-10 & -110.6 & -502.9 & 893--932, 503--539 & 100.6 \\
59 & 2014-01-10 & 538.3 & -369.1 & 1545--1577, 639--677 & 61.8 \\
60 & 2014-01-10 & 507.7 & -352.4 & 1524--1541, 663--681 & 43.6 \\
61 & 2014-01-10 & 334.3 & 73.6 & 1350--1367, 1089--1107 & 70.5 \\
62 & 2014-01-10 & -159.0 & 388.5 & 857--872, 1406--1420 & 32.3 \\
63 & 2014-01-10 & -617.3 & 70.5 & 401--413, 1088--1101 & 59.1 \\
64 & 2014-01-11 & 87.7 & 382.4 & 1105--1118, 1400--1413 & 67.0 \\
65 & 2014-01-11 & 534.4 & 81.9 & 1550--1567, 1098--1115 & 37.8 \\
66 & 2014-01-11 & -565.4 & -383.3 & 450--470, 633--654 & 55.0 \\
67 & 2014-01-11 & -532.7 & -331.1 & 485--498, 685--699 & 49.6 \\
68 & 2014-01-11 & 472.7 & -508.4 & 1473--1546, 486--557 & 65.9 \\
69 & 2014-01-11 & -674.0 & 226.7 & 336--364, 1238--1266 & 41.8 \\
70 & 2014-01-11 & -161.1 & -244.3 & 854--871, 772--787 & 56.6 \\
71 & 2014-01-11 & 6.6 & -519.8 & 1019--1043, 489--519 & 34.6 \\
72 & 2014-01-11 & -474.2 & -542.6 & 541--557, 473--489 & 48.4 \\
73 & 2014-01-12 & -62.2 & -328.4 & 949--975, 684--708 & 60.0 \\
74 & 2014-01-12 & -179.7 & -411.4 & 834--855, 602--622 & 82.1 \\
75 & 2014-01-12 & -10.2 & -271.6 & 1008--1020, 746--759 & 126.0 \\
76 & 2014-01-12 & 608.3 & -356.0 & 1625--1640, 662--675 & 55.0 \\
77 & 2014-01-12 & 608.9 & -382.4 & 1627--1639, 635--648 & 55.0 \\
78 & 2014-01-12 & 622.5 & -271.9 & 1625--1668, 731--775 & 67.0 \\
79 & 2014-01-12 & 196.2 & -508.0 & 1202--1237, 501--531 & 73.1 \\
80 & 2014-01-12 & -639.8 & -465.0 & 377--392, 552--565 & 91.4 \\
81 & 2014-01-12 & -477.8 & -434.5 & 539--553, 582--596 & 82.1 \\
\textcolor{red}{82} & \textcolor{red}{2014-01-13} & \textcolor{red}{67.2} & \textcolor{red}{-381.7} & \textcolor{red}{1074--1113, 622--661} & \textcolor{red}{85.6} \\
83 & 2014-01-13 & 108.3 & -356.7 & 1126--1140, 661--675 & 85.6 \\
84 & 2014-01-13 & 184.8 & -272.3 & 1193--1225, 737--768 & 80.4 \\
85 & 2014-01-13 & 109.3 & -357.0 & 1125--1143, 659--676 & 80.4 \\
86 & 2014-01-13 & 714.5 & -357.5 & 1732--1745, 661--673 & 45.6 \\
87 & 2014-01-14 & 366.6 & -259.9 & 1383--1398, 757--773 & 51.5 \\
88 & 2014-01-14 & -699.7 & -102.7 & 308--340, 904--934 & 60.0 \\
89 & 2014-01-14 & -187.0 & 117.4 & 824--855, 1125--1161 & 43.6 \\
90 & 2014-01-14 & -132.6 & 117.9 & 884--900, 1134--1151 & 54.3 \\
91 & 2014-01-14 & 583.0 & 385.8 & 1589--1640, 1399--1423 & 63.8 \\
\end{longtable}
}

\section{Conclusions}
In this work, we have developed a robust automatic pipeline for the systematic detection of oscillatory events in GONG \ha observations. The framework integrates automated filament segmentation with a multi-scale spectral analysis, where spatially degraded signals are evaluated using Lomb--Scargle periodograms, CNN-driven background estimation, and an empirical conformal calibration of significance thresholds. By consolidating detections across both period and space, the pipeline ensures that only oscillatory events validated by a rigorous cross-scale consistency criterion are included in the final catalog.

The pipeline's primary objective is the recovery of spatially coherent, filament-scale oscillatory behavior by mitigating the localized variability that characterized previous pixel-wise analyses \citep{luna2022, castello2025}.
Central to this approach is the multi-scale averaging strategy, where larger kernels emphasize the collective dynamics of extended filament regions, while the cross-scale consistency criterion effectively filters out spurious or isolated detections.
An a posteriori analysis justifies this approach, showing that SNR is maximized when the averaging scale matches the spatial coherence of the filament motion.

Consequently, the framework provides a scalable and automated solution for identifying robust global oscillation modes with high statistical confidence.
The validation over the first two weeks of January 2014 confirms the pipeline's effectiveness, accurately recovering multiple events from the manual catalog of \citet{luna_gong_2018} with spatial and spectral agreement. The consistency between our automated detections and traditional slit-based analysis, as seen in the January 1st example, reinforces the reliability of the method. Crucially, the pipeline identifies a significantly higher volume of events than manual surveys, demonstrating the superior sensitivity gained through a systematic search across all filament regions and scales.
This significant increase in detections demonstrates that our framework surpasses the sensitivity of traditional visual inspection, establishing it as a superior tool for large-scale statistical investigations. Of particular importance is the successful recovery of oscillatory events on days and in regions where \citet{luna_gong_2018} reported no activity. The case study of January 13th, 2014, confirms that these additional detections represent genuine oscillatory signatures--verifiable through conventional time-distance analysis--thereby substantiating the physical validity and scientific value of the automated catalog.

The comparative analysis also provides important insights into the nature of the observed discrepancies. Detections are excluded when they fail to meet our rigorous criteria, whether due to missing initial segmentation, signals that do not reach our significance thresholds, or oscillations that lack sufficient cross-scale persistence. Furthermore, instances of differing periods arise from the contrast between localized slit measurements and our scale-integrated, spatially coherent detection strategy. These differences reflect the distinct ways in which manual slit-based analysis and automated spectral detection capture the global oscillatory behavior of filaments, each with its own observational perspective and sensitivity.

A limitation of the current implementation is that the Lomb--Scargle analysis provides a global, time-integrated estimate of the dominant period within the selected interval. The pipeline does not require the signal to be strictly stationary; it can accommodate damped, growing, or amplitude-modulated oscillations, provided the dominant period remains sufficiently coherent over the analyzed temporal and spatial windows. However, the method is not designed to track instantaneous frequency variations. Strongly non-stationary signals, such as very short wave packets or pronounced frequency drifts, may therefore fail to meet the significance criteria or assigned a time-averaged period. Such cases are better suited for dedicated time-frequency methods like wavelet analysis. The pipeline should therefore be considered as a robust tool for the automated discovery of spatially coherent dominant periodicities in large datasets, rather than for the characterization of rapidly evolving frequencies.

The proposed framework provides a robust foundation for the systematic study of filament oscillations across long-term GONG \ha datasets. Its primary strength lies in transitioning from manual, low-throughput procedures to a reproducible and high-performance pipeline. With a total processing time of approximately 80 minutes per observing day, the pipeline enables the efficient construction of massive oscillation catalogs over extended time intervals. This capability paves the way for studies spanning the entire solar cycle, providing the data necessary to investigate occurrence rates, triggering mechanisms, and the physical properties of filaments in relation to their long-term solar evolution. In a forthcoming study, Paper~II, we will analyze the properties of these oscillations over at least one full solar cycle.

The present work should be regarded as a methods paper and a first validation of the pipeline, providing the basis for future large-scale statistical studies of filament oscillations. While several improvements are possible, the current implementation uses conservative criteria---such as first-frame filament detection and strict multi-scale support---to ensure the reliability of the detected events. Although this version focuses on identifying the location and period of each oscillation, the extraction of onset times, amplitudes, polarization, and damping rates is planned for a subsequent study in this series. This expansion will allow for a more comprehensive physical characterization and seismological analysis of the detected events, building upon the detection framework established here.

\section*{Code availability}
The code is publicly available in the following public repository:
\url{https://github.com/GuillemCastello/SolFilOsc}

\begin{acknowledgements}
This publication is part of the I+D+i project PID2023-147708NB-I00 funded by MICIU/AEI/10.13039/501100011033 and by  FEDER, EU. G. Castelló acknowledges financial support from the Direcció General d'Universitats, Recerca i Ensenyaments Artístics Superiors of the Government of the Balearic Islands through a pre-doctoral fellowship co-financed by the European Social Fund Plus (FSE+) within the framework of the Balearic Islands Programme 2021–2027. Co-funded by the European Union.
\end{acknowledgements}

%
\bibliographystyle{aa.bst} 
\bibliography{bibliography.bib} 
%

\begin{appendix}
\nolinenumbers
\section{Preprocessing visualization}\label{appendix:preprocessing-visualization}

\begin{figure*}[!ht]
\centering
\includegraphics[width=\linewidth]{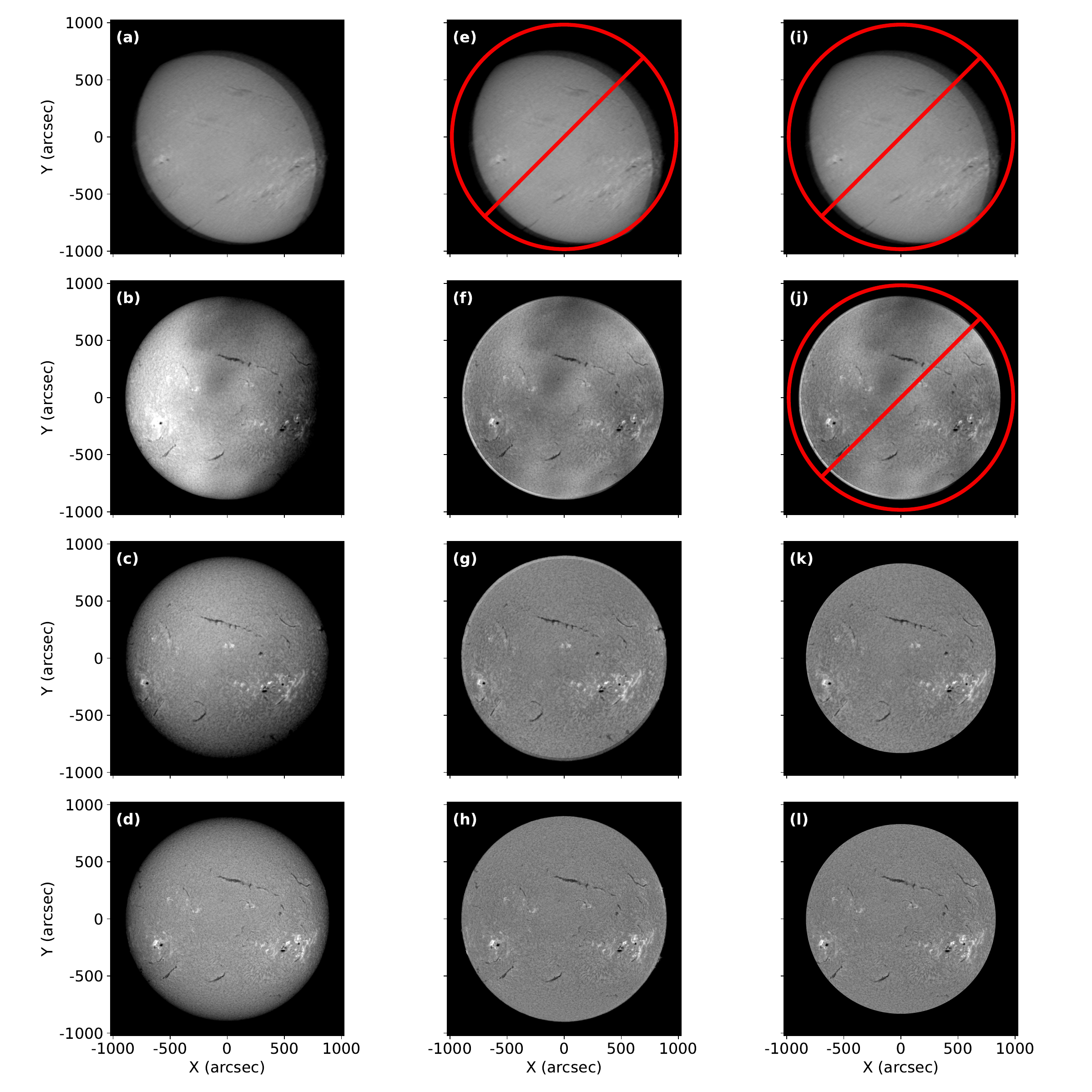}
\caption{Preprocessing pipeline stages for representative GONG H$\alpha$ frames. Rows display examples with different initial quality levels, ranging from very low (top) to good (bottom). Left column: Raw input frames from the GONG network. Middle column: Intermediate results after initial selection, limb-darkening correction, and spatial background correction. Right column: Final preprocessed frames following outlier detection, temporal alignment to 12:00:00 UT, and solar disk masking. A red crossed-circle icon denotes frames discarded by the pipeline at the corresponding stage.}

\label{fig:preprocessing_steps}
\end{figure*}

To illustrate the data preprocessing pipeline described in Section \ref{sec:data-preprocessing},
Figure \ref{fig:preprocessing_steps} provides a qualitative visualization of the procedure applied to four representative GONG H$\alpha$ frames.

The rows are organized by initial image quality, ranging from very low (top) to good quality (bottom). Each column represents a successive stage of the pipeline: the raw GONG observations (left), the intermediate results after initial selection and limb-darkening/spatial background corrections (middle), and the final output following outlier detection, temporal alignment to 12:00:00 UT, and solar disk masking (right). Frames that fail to meet the quality thresholds at any stage are marked with a red crossed-circle icon.

The very low-quality frame \cref{fig:preprocessing_steps}(a) is discarded at the initial filtering stage \cref{fig:preprocessing_steps}(e,i). The low-quality frame \cref{fig:preprocessing_steps}(b) passes the first selection but is ultimately rejected during the outlier-detection step \cref{fig:preprocessing_steps}(j) due to persistent cloud contamination. In contrast, the medium- and good-quality examples \cref{fig:preprocessing_steps}(c,d) pass all stages and are retained for analysis. Specifically, the transition from the raw frames to the corrected versions \cref{fig:preprocessing_steps}(g,h) highlights the effectiveness of the limb-darkening and background corrections in mitigating large-scale intensity gradients, such as the darkening on the eastern limb. This process ensures the spatial homogeneity and temporal consistency required for the subsequent segmentation and spectral analysis, as seen in the final processed frames \cref{fig:preprocessing_steps}(k,l).

\section{Conformal calibration of the spectral detection threshold}
\label{appendix:cp}

A pixel-level intensity time series observed by a solar telescope can be written schematically as
\begin{equation}
    I(t) = I_{\mathrm{signal}}(t) + \epsilon(t),
\end{equation}
where \(I_{\mathrm{signal}}(t)\) denotes the component of scientific interest, in our case a possible periodic signal, and \(\epsilon(t)\) represents the remaining stochastic and instrumental contributions to the observation.

For each time series, we compute a Lomb--Scargle or generalized periodogram, denoted by \(\mathcal{P}(f)\). We interpret this periodogram as a noisy estimator of an underlying power spectral density (PSD),
\begin{equation}
    \mathcal{P}(f) = \mathcal{S}(f)\,\epsilon_{\mathrm{aleatoric}}
    =
    \left[\mathcal{S}_{\mathrm{signal}}(f) + \mathcal{S}_{\mathrm{noise}}(f)\right]\epsilon_{\mathrm{aleatoric}},
\end{equation}
where \(\mathcal{S}(f)\) is the true PSD, \(\mathcal{S}_{\mathrm{signal}}(f)\) is the contribution associated with the physical signal of interest, \(\mathcal{S}_{\mathrm{noise}}(f)\) is the background noise PSD, and \(\epsilon_{\mathrm{aleatoric}}\) represents the intrinsic stochastic variability of the periodogram estimator. In the absence of a coherent signal, \(I_{\mathrm{signal}}(t)=0\), the time series is purely noise and the observed periodogram fluctuates around the noise background,
\begin{equation}
    \mathcal{P}(f) = \mathcal{S}_{\mathrm{noise}}(f)\,\epsilon_{\mathrm{aleatoric}}.
\end{equation}

In solar observations such as those analyzed here, background power spectra are commonly described by a smooth power-law component, sometimes with an additive white-noise floor at high frequencies. In this work, the adopted background model is the parametric form given in Eq.~\eqref{eq:noise_model}. The role of the CNN introduced in \citet{castello2025} is to estimate the parameters of this background model from an observed periodogram,
\begin{equation}
    \mathrm{CNN}\!\left(\mathcal{P}(f)\right)
    \rightarrow
    \hat{\Theta} = \{\hat{a}, \hat{\alpha}, \hat{b}\}
    \rightarrow
    \hat{\mathcal{S}}(f;\hat{\Theta}),
\end{equation}
where \(\hat{\mathcal{S}}(f;\hat{\Theta})\) denotes the CNN-estimated background PSD. This estimate is not assumed to be exact. It contains both the uncertainty intrinsic to the periodogram and the approximation error of the learned model.

We therefore calibrate the detection threshold using a conformal-prediction strategy. Conformal prediction provides a distribution-free framework for converting calibration residuals into finite-sample uncertainty sets under exchangeability assumptions \citep{Vovk2005}. In the present setting, we use this idea not to construct prediction intervals for a scalar response, but to calibrate a frequency-dependent upper threshold for periodogram power.

The calibration set is constructed from pixels sampled outside the filament regions. Over the period range considered in this work, from 3 hours down to 16.66 minutes, these pixels are assumed not to contain the filament oscillations targeted by the pipeline. For a calibration pixel, the observed time series is therefore modeled as
\begin{equation}
    I_{\mathrm{cal}}(t) = \epsilon(t),
\end{equation}
and its periodogram can be written as
\begin{equation}
    \mathcal{P}_{\mathrm{cal}}(f)
    =
    \mathcal{S}_{\mathrm{cal}}(f)\,\epsilon_{\mathrm{aleatoric}}.
\end{equation}
Passing this calibration periodogram through the CNN gives
\begin{equation}
    \mathrm{CNN}\!\left(\mathcal{P}_{\mathrm{cal}}(f)\right)
    \rightarrow
    \hat{\Theta}_{\mathrm{cal}}
    =
    \{\hat{a},\hat{\alpha},\hat{b}\}
    \rightarrow
    \hat{\mathcal{S}}_{\mathrm{cal}}(f;\hat{\Theta}_{\mathrm{cal}}).
\end{equation}

We compute calibration residuals in logarithmic space. This is convenient because the dominant periodogram variability is multiplicative in linear power, whereas the logarithmic transform makes it additive:
\begin{equation}
    \log \mathcal{P}_{\mathrm{cal}}(f)
    =
    \log \mathcal{S}_{\mathrm{cal}}(f)
    +
    \log \epsilon_{\mathrm{aleatoric}}.
\end{equation}
The residual score for a calibration pixel is then defined as
\begin{multline}
    \mathcal{R}_{\mathrm{cal}}(f)
    =
    \log \mathcal{P}_{\mathrm{cal}}(f)
    -
    \log \hat{\mathcal{S}}_{\mathrm{cal}}(f;\hat{\Theta}_{\mathrm{cal}})
    \\
    =
    \left[
    \log \mathcal{S}_{\mathrm{cal}}(f)
    -
    \log \hat{\mathcal{S}}_{\mathrm{cal}}(f;\hat{\Theta}_{\mathrm{cal}})
    \right]
    +
    \log \epsilon_{\mathrm{aleatoric}}.
\end{multline}
The first term represents the error of the CNN background estimate, while the second term represents the irreducible variability of the periodogram estimator. Thus, the residual distribution estimated from quiet-Sun pixels captures both sources of uncertainty under the null hypothesis of no target oscillatory signal.

Repeating this procedure for all \(N_{\mathrm{cal}}\) calibration pixels gives a set of residual curves
\begin{equation}
    \left\{
    \mathcal{R}_i(f)
    \right\}_{i=1}^{N_{\mathrm{cal}}}.
\end{equation}
In the implementation used in this work, these residuals are robustly standardized at each frequency using the median and median absolute deviation. This step removes frequency-dependent bias and scale variations, so that the conformal scores are calibrated on a comparable scale across the spectrum. Similar score-normalization ideas are commonly used in locally adaptive conformal methods, where residuals are rescaled to account for non-constant uncertainty \citep{lei2018,Romano2019}. For clarity of exposition, the equations below are written for the unstandardized residuals \(\mathcal{R}_i(f)\), but the same construction is applied to the standardized scores \(\mathcal{Z}_i(f)\) in the pipeline.
For a target miscoverage level \(\delta\), we compute a frequency-dependent empirical conformal quantile. Let
\begin{equation}
    k =
    \left\lceil
    (N_{\mathrm{cal}}+1)(1-\delta)
    \right\rceil .
\end{equation}
Then, for each frequency \(f\), we define
\begin{equation}
    q_{1-\delta}(f)
    =
    \mathcal{R}_{(k)}(f),
\end{equation}
where \(\mathcal{R}_{(k)}(f)\) is the \(k\)-th order statistic of the calibration residuals at frequency \(f\). This is the standard split-conformal finite-sample correction, using \((N_{\mathrm{cal}}+1)(1-\delta)\) rather than the naive empirical \((1-\delta)\)-quantile \citep{lei2018}. In practice, this requires \(k \leq N_{\mathrm{cal}}\); otherwise, the requested confidence level is more extreme than what can be resolved by the finite calibration sample.
The quantity \(q_{1-\delta}(f)\) is local in frequency. It estimates, under the quiet-Sun calibration distribution, how far above the CNN-predicted background a periodogram value can fluctuate at frequency \(f\) due only to background-model error and the intrinsic variability of the periodogram estimator. Therefore, if the CNN has frequency-dependent errors, or if some frequency ranges are intrinsically more variable than others, this behavior is incorporated into the calibrated threshold.
For a test pixel, which may or may not contain an oscillatory signal, we first estimate its background PSD using the CNN:
\begin{equation}
    \mathrm{CNN}\!\left(\mathcal{P}_{\mathrm{test}}(f)\right)
    \rightarrow
    \hat{\mathcal{S}}_{\mathrm{test}}(f).
\end{equation}
The calibrated upper threshold in linear power is then
\begin{equation}
    \tau(f)
    =
    \hat{\mathcal{S}}_{\mathrm{test}}(f)
    \exp\!\left(q_{1-\delta}(f)\right).
\end{equation}
A periodogram peak satisfying
\begin{equation}
    \mathcal{P}_{\mathrm{test}}(f) > \tau(f)
\end{equation}
is therefore interpreted as exceeding the level expected from the calibrated quiet-Sun background at frequency \(f\).
This construction should be understood as a pointwise, frequency-dependent conformal threshold. It calibrates the exceedance probability independently at each frequency, rather than providing a simultaneous guarantee over the full spectral curve. Conformal methods for functional or curve-valued outputs provide the appropriate framework when simultaneous bands over an entire function domain are required \citep{Diquigiovanni2022}. In the present work, the pointwise formulation is sufficient because candidate oscillatory events are subsequently consolidated through spectral, spatial, and multi-scale consistency criteria.

\subsection{Validation of significance levels using MCMC}

\begin{figure}[!ht]
\centering
\includegraphics[width=\linewidth]{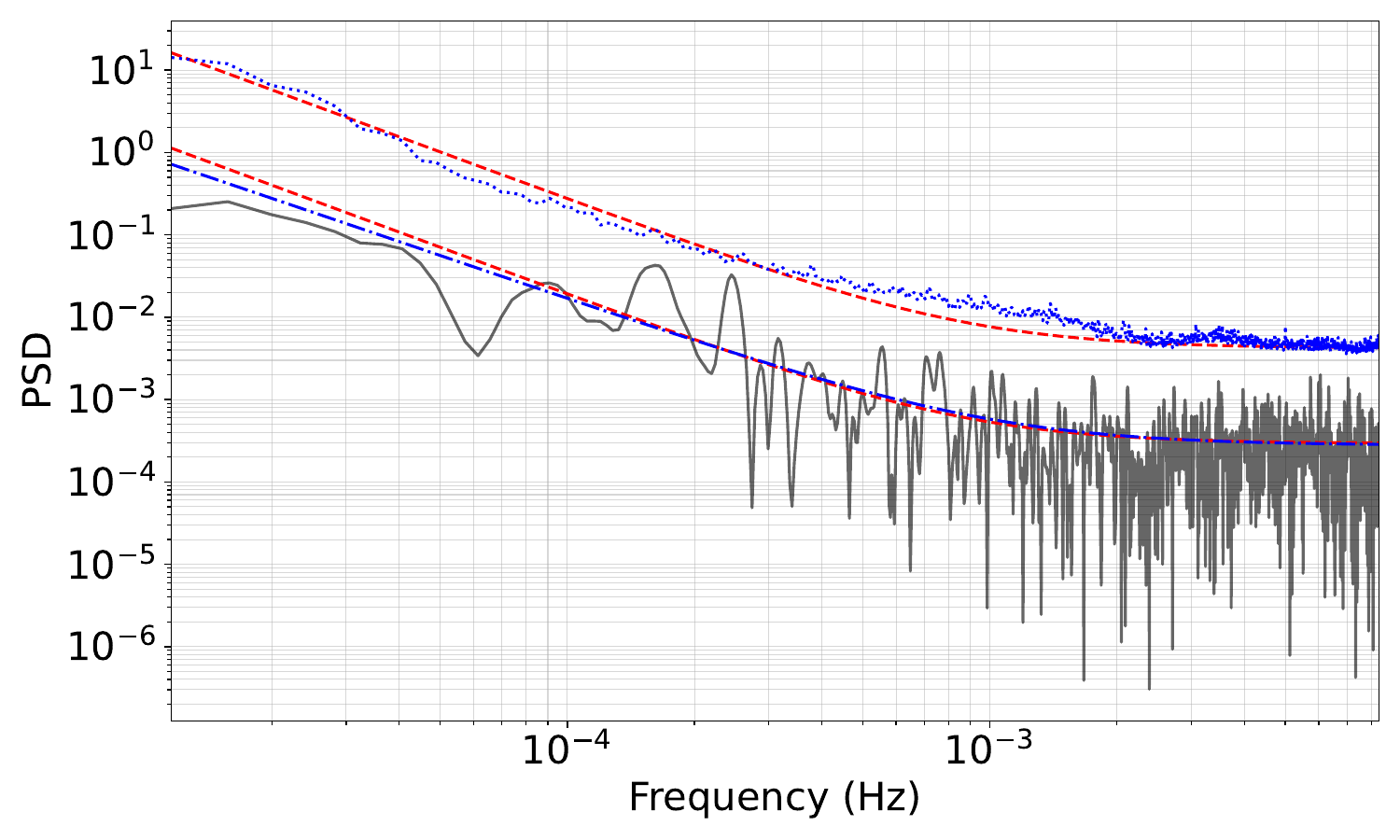}
\caption{PSD of a representative on-disk pixel from November 22, 2014. Best-fit and $5\sigma$ confidence lines obtained with the MCMC analysis are shown in red, while those obtained with the CNN+CP framework are shown in blue.}
\label{fig:comparison_mcmc_cnn}
\end{figure}

The CP framework is validated by comparing its results against a standard Markov Chain Monte Carlo (MCMC) analysis of the power spectral density (PSD). Figure~\ref{fig:comparison_mcmc_cnn} illustrates the PSD of a representative on-disk pixel, showcasing the background best-fit and $5\sigma$ confidence thresholds derived from both methodologies.

The best-fit curves are, as expected from the results reported in \citet{castello2025}, very similar. This confirms that the CNN provides an accurate approximation to the MCMC-derived best-fit background model. 
Similarly, the confidence thresholds show a high degree of overlap across the frequency range.
 
The small observed deviations reflect the different nature of the two approaches. MCMC thresholds depend on explicit parametric assumptions and prior distributions, where poorly specified priors can bias the posterior distribution or hinder convergence, which may directly affect the inferred confidence thresholds.
In contrast, the CP-based thresholds are derived empirically from day-wise calibration residuals.
This empirical approach allows the CP framework to naturally incorporate daily systematic effects or instrumental noise that might not be captured by a fixed parametric model.

The primary advantage of the CNN+CP pipeline is its computational efficiency. While MCMC analysis requires several minutes per pixel, the CNN+CP framework processes the entire solar disk in milliseconds per pixel after a brief ($\sim$ 10 min) daily calibration step. As demonstrated, this gain in speed is achieved with negligible loss in statistical accuracy, providing a robust and scalable solution for large-scale GONG observations.

\end{appendix}
\end{document}